\documentclass[12pt]{article}
\usepackage{amsmath,amsthm,amsfonts}
\usepackage{url}
\newcommand{\pf}{\noindent {\bf Proof:} }
\DeclareMathAlphabet{\mathpzc}{OT1}{pzc}{m}{it}
\newtheorem{theorem}{Theorem}
\newtheorem{remark}{Remark}

\newtheorem{corollary}[theorem]{Corollary}
\newtheorem{lemma}[theorem]{Lemma}

\newtheorem{conjecture}[theorem]{Conjecture}

\title{Lexicographically least words in the orbit closure of the Rudin-Shapiro word}
\author{James Currie\thanks{The author is
supported by an NSERC Discovery Grant.}\\
Department of Mathematics and Statistics \\
University of Winnipeg \\
515 Portage Avenue \\
Winnipeg, Manitoba R3B 2E9 (Canada) \\
\url{j.currie@uwinnipeg.ca} }
\begin{document}
\date{\today}
\maketitle
\begin{abstract}
\noindent \title{We give an effective characterization of the lexicographically least word in the orbit closure of the Rudin-Shapiro word ${\bf w}$ having a specified prefix. In particular, the lexicographically least word in the orbit closure of the Rudin-Shapiro word is $0{\bf w}$. This answers a question Allouche et al.}
\vspace{.1in}
\end{abstract}
\section{Introduction} Let $f:\{a,b,c,d\}^*\rightarrow\{0,1\}^*$ and $g:\{a,b,c,d\}^*\rightarrow\{a,b,c,d\}^*$
be given respectively by
\begin{eqnarray*}
f(a)&=&0\\
f(b)&=&0\\
f(c)&=&1\\
f(d)&=&1
\end{eqnarray*}
and
\begin{eqnarray*}
g(a)&=&ab\\
g(b)&=&ac\\
g(c)&=&db\\
g(d)&=&dc.
\end{eqnarray*}

\noindent Let ${\bf u}=g^\omega(a)$. The {\bf Rudin-Shapiro word $w$} is given by ${\bf w}=f({\bf u})$.
Thus
$${\bf w} = 00010010000111\cdots$$

The Rudin-Shapiro word has been the subject of much study  in
combinatorics on words. A standard reference is \cite{AS}. An
alternative characterization of the Rudin-Shapiro word is as
follows: For each non-negative integer $n$, let $P(n)$ denote the
parity of the number of times 11 appears in the binary
representation of $n$. For example, 59 has binary representation
111011,  which contains 3 occurrences of 11, so that
$P(59)=1\equiv 3$ (mod 2). The Rudin-Shapiro word is the infinite
binary word whose $i^{th}$ bit (starting at $i=0$ on the left) is
$P(i)$.

\begin{remark}\label{0w} From this second characterization, it follows that
if $p$ is any finite prefix of ${\bf w}$, then $0p$ is a factor of
${\bf w}$; indeed, choose odd $s> |p|$. Then  the binary
representation of $2^s-1$ is a string of $1$'s of length $s$,
whence $P(2^s-1)=0$. On the other hand, $P(i)=P(2^s+i)$ for $0\le
i\le |p|-1$, so that $0p$ appears in ${\bf w}$, starting at bit
$2^s$.
\end{remark}

We will freely use standard notions from combinatorics on words.
For further information see \cite{lothaire}, for example. Let the
set of finite factors of ${\bf u}$ be denoted by ${\mathpzc F}$.
Any word $v\in{\mathpzc F}$ is a factor of a word $g(v'),
v'\in{\mathpzc F}$ where $|v'|\le \frac{|v|+2}{2}.$ Since
$\frac{|v|+2}{2}<|v|$ if $|v|>2$ and $\epsilon, 0, 1, 00, 01, 10,
11\in {\mathpzc F}$, it follows that we can determine
membership/nonmembership in ${\mathpzc F}$ effectively. We can
also therefore  determine effectively whether or not a word is a
factor of $\bf w$.

The {\bf orbit closure} of a right infinite word ${\bf v}$ is the
set of those right infinite words whose every finite prefix is a
factor of ${\bf v}$. We denote the orbit closure of $\bf v$ by
${\mathpzc O}_v$. Remark~\ref{0w} shows that $0{\bf w}$ is in
${\mathpzc O}_w$. Consider the natural order on $\{0,1\}^*$,
namely the lexicographic order generated by  $0<1$. In this
article we will say simply `least' for `lexicographically least'
with respect to this order.

Recently it was conjectured \cite{ARS} that
\begin{conjecture}\label{jeff}
Word $0{\bf w}$ is the least word
in ${\mathpzc O}_w$.
\end{conjecture}
 We prove and generalize this conjecture.

\section{Least words in ${\mathpzc O}_u$.}
We begin this section with a few observations and notations.
Consider the natural order on $\{a,b,c,d\}^*$, namely the
lexicographic order generated by  $a<b<c<d$. Under this order,
morphism $g$ is strictly order-preserving; i.e., $g(x)< g(y)$ if
$x< y$. If $w$ is a word with prefix (resp., suffix) $v$, we will
use the notation $v^{-1}w$ (resp., $wv^{-1}$) to denote the word
obtained from $w$ by erasing the prefix (resp., suffix) $v$. Thus
writing $w=vu$, we have $v^{-1}w=u$. We say that word $x$ appears
with {\bf index} $i$ in word $y$ if we can write $y=pxq$ for some
words $p$ and $q$ where $|p|=i$.
\begin{remark}\label{index}
Letters $a$ and $d$ only ever appear in ${\bf u}$ with even index,
while letters $b$ and $c$ only ever appear in ${\bf u}$ with odd
index.
\end{remark}
\begin{lemma}\label{pullback} Let $\mu\in\mathpzc F$, $|\mu|\ge 3$.
We can uniquely choose words $x\in\{\epsilon, a, d\}$, $y\in\{\epsilon, b, c\}$ and $\hat{\mu}\in \mathpzc F$ such that $x\mu y=g(\hat{\mu})$. Further, $|\hat{\mu}|<|\mu|$.\end{lemma}
\pf The existence of such $x$, $y$ and $\mu$ is clear; we shall establish uniqueness. It will suffice to show the uniqueness of $x$ and $y$ since $g$ is strictly order preserving, hence   invertible. Let us give the proof that $x$ is unique:

If $a$ or $d$ is a prefix of $\mu=x^{-1}g(\hat{\mu})y^{-1}$, then clearly $x$ must be $\epsilon.$

Suppose that $b$ is a prefix of $\mu$.
The length 4 elements of $\mathpzc F$ having $b$ as a second letter are
\begin{eqnarray*}
g(ab)&=&abac\\
g(ac)&=&abdb\\
g(ca)&=&dbab\\
g(cd)&=&dbdc.
\end{eqnarray*}

These elements have distinct length 3 suffixes. If $b$ is a prefix of $\mu$, then since $|\mu|\ge 3$, let $\mu'$ be the length 3 prefix of $\mu$. We see that $x\mu'$ must be one of the length 4 factors on the above list, and hence,  the length 3 prefix of $\mu$ determines whether $x$ is $a$ or $d$; thus, $x$ is uniquely determined if $b$ is a prefix of $\mu$.

A similar argument dispatches the case where $\mu$ starts with a $c$, so that $x$ is uniquely determined in all cases. An analogous argument shows that $y$ is uniquely determined.

Finally, we have $2|\hat{\mu}|=|g(\hat{\mu})|=|x\mu y|\le
|\mu|+2$. Then, $|\hat{\mu}|\le\frac{|\mu|+2}{2}<|\mu|$ since $2<
|\mu|.\Box$\vspace{.1in}

\begin{lemma}\label{mu} Let $\mu\in\mathpzc F$, $|\mu|\ge 3$.
Write $x\mu y=g(\hat{\mu})$, where $x\in\{\epsilon, a, d\}$, $y\in\{\epsilon, b, c\}$ and $\hat{\mu}\in \mathpzc F$. The least word in ${\mathpzc O}_u$ having prefix $\mu$ is $x^{-1}g(\nu)$, where $\nu$ is the least word in ${\mathpzc O}_u$ having prefix $\hat{\mu}$
\end{lemma}
\pf Let $\nu$ be the least word in ${\mathpzc O}_u$ having prefix
$\hat{\mu}$. Let $\lambda$ be any word in ${\mathpzc O}_u$ having
prefix $\mu$. Let ${\mathpzc l}$ be the prefix of $\lambda$ of
length $g(\hat{\mu})-|x|$. Since $\mu$ is a prefix of ${\mathpzc
l}$ it follows from Lemma~\ref{pullback} that $x{\mathpzc
l}=g(\hat{\mu})$, and we may write $x\lambda = g(\hat{\lambda})$,
where $\hat{\lambda}$ is a word in ${\mathpzc O}_u$ with prefix
$\hat{\mu}$. Since $g$ is order preserving, the result
follows.$\Box$\vspace{.1in}

When $|\mu|\ge 3$, finding the least word in ${\mathpzc O}_u$ with prefix $\mu$ reduces to solving the same problem for a shorter word. It remains to give the least words with specified short prefixes.
\begin{lemma}\label{u}The least word in ${\mathpzc O}_u$ is $\bf u$.
\end{lemma}
\pf Let $\lambda$ be the least word in ${\mathpzc O}_u$. The least
length 3 word in $\mathpzc F$ is $aba$. Since $\bf u$ has $aba$ as
a prefix, $\lambda$ must also have $aba$ as a prefix. Letting $\mu
= aba$ and applying Lemma~\ref{mu}, we find $x = \epsilon$,
$\hat{\mu} = ab$, and $\nu$ is the least word of ${\mathpzc O}_u$
with prefix $ab$. This implies $\nu = \lambda$, whence $\lambda =
x^{-1}g(\nu)=g(\lambda)$. Then $\lambda$ is the fixed point of $g$
with first letter $a$, namely $\bf u.\Box$\vspace{.1in}

 \begin{corollary} If $\mu{\bf u}$ is in ${\mathpzc O}_u$, then $\mu {\bf u}$ is the least word with prefix $\mu$ in ${\mathpzc O}_u$.
 \end{corollary}
 \pf If $\lambda$ is any word with prefix $\mu$ in ${\mathpzc O}_u$, write $\lambda = \mu\lambda'$. Since $\lambda'\in {\mathpzc O}_u,$ by the previous lemma, $\lambda'$ cannot be lexicographically less than $\bf u$. It follows that $\lambda$ cannot be lexicographically less than $\mu {\bf u}.\Box$\vspace{.1in}
 \begin{corollary}\label{prefix} Suppose that $\mu{\bf u}$ is in ${\mathpzc O}_u$, $\pi$ is a prefix of $\mu {\bf u}$, and $\mu$ is a prefix of $\pi$. Then $\mu{\bf u}$  is the least word with prefix $\pi$ in ${\mathpzc O}_u$.
 \end{corollary}
 \pf Every word of ${\mathpzc O}_u$ with prefix $\pi$ has prefix $\mu$, and by the last lemma must be lexicographically at least as great as $\mu {\bf u}.\Box$\vspace{.1in}

 \begin{lemma}\label{bu cu} Words $b{\bf u}$, $c{\bf u}$ are in ${\mathpzc O}_u.$
 \end{lemma}
 \pf Every factor of $\bf u$ appears in $\bf u$ infinitely often. Let $\mu$ be any prefix of $\bf u$. Then  $\mu$ is a prefix of $g(\mu)$; also, the first letter of $\mu$ is $a$. The second occurrence of $\mu$ in $\bf u$ thus occurs either in the context $b\mu$ or $c\mu$. However, if $b\mu$ is a factor of $\bf u$, then so is $g(b\mu)=acg(\mu)$, which contains $c\mu$ as a factor.
Similarly, if $c\mu$ is a factor of $\bf u$, then so is
$g(c\mu)=dbg(\mu)$, which contains $b\mu$ as a factor.  We
conclude then, that both $b\mu$ and $c\mu$ are factors of $\bf u$.
Since $\mu$ was an arbitrary prefix of $\bf u$, we conclude that
$b{\bf u}$ and $c{\bf u}$ are in  ${\mathpzc
O}_u.\Box$\vspace{.1in}

\begin{corollary}\label{g^n}For non-negative integers $n$, words $g^n(b){\bf u}$, $g^n(c){\bf u}$ are in ${\mathpzc O}_u.$ In particular, words $ac{\bf u}$, $db{\bf u}$ , $abdb{\bf u}$, $dcac{\bf u}$, $dcdbabdb{\bf u}$ are in ${\mathpzc O}_u.$
\end{corollary}

\begin{lemma}\label{table} For given $\mu$, $|\mu|\le 2$, the least word in ${\mathpzc O}_u$ with prefix $\mu$ is as given in the following table:\vspace{.1in}

\begin{tabular}{|c|c|}\hline
$\mu$&least word\\\hline
$\epsilon$, $a$, $ab$&${\bf u}$\\\hline
$ac$&$ac{\bf u}$\\\hline
$b$, $ba$&$b{\bf u}$\\\hline
$bd$&$bdb{\bf u}$\\\hline
$c$, $ca$&$c{\bf u}$\\\hline
$cd$&$cdbabdb{\bf u}$\\\hline
$d$, $db$&$db{\bf u}$\\\hline
$dc$&$dcac{\bf u}$\\\hline
\end{tabular}
\end{lemma}
\pf The least words listed for $\epsilon$, $a$, $ab$, $ac$, $b$, $ba$, $c$, $ca$, $db$ and $dcac$ follow immediately from Corollary~\ref{prefix} and Corollary~\ref{g^n}. This leaves us to establish the correctness of the words listed for prefixes $d$, $bd$ and $cd$.
\subsubsection*{Prefix $d$} The least length 2 word in ${\mathpzc F}$ starting with $d$ is $db$. It follows that the least word in ${\mathpzc O}_u$ with prefix $d$ will be the least word in ${\mathpzc O}_u$ with prefix $db$, namely $db{\bf u}$.
\subsubsection*{Prefix $bd$} The least length 3 word in ${\mathpzc F}$ starting with $bd$ is $bdb$. It follows that the least word in ${\mathpzc O}_u$ with prefix $bd$ will be the least word in ${\mathpzc O}_u$ with prefix $bdb$, namely $bdb{\bf u}$.
\subsubsection*{Prefix $cd$} The least length 5 word in ${\mathpzc F}$ starting with $cd$ is $cdbab$. It follows that the least word in ${\mathpzc O}_u$ with prefix $d$ will be the least word in ${\mathpzc O}_u$ with prefix $cdbab$. As per Lemma~\ref{mu}, this word will be $d^{-1}g(\nu)$, where $\nu$ is the least word in ${\mathpzc O}_u$ with prefix $dca$. This means that $\nu = dcac{\bf u}$. Then $d^{-1}g(\nu)=cdbabdb{\bf u}$, as desired.$\Box$\vspace{.1in}
\begin{theorem}\label{Ou} Let $\mu\in{\mathpzc F}$. The least word in ${\mathpzc O}_u$ with prefix $\mu$ can be effectively determined, and has the form $\mu'{\bf u}$ for some $\mu'\in{\mathpzc F}$.
\end{theorem}
\pf This follows by combining Lemma~\ref{mu} and Lemma~\ref{table}.
\section{Least words in ${\mathpzc O}_w$}
We begin with a few observations and notations. Morphism $f$ is
order-preserving; i.e., $f(x)\le f(y)$ if $x\le y$.  Say that
factor $v$ of ${\bf w}$ is {\bf ambiguous} if it occurs in ${\bf
w}$ sometimes with even index, and sometimes with odd index.
Evidently, every factor of an ambiguous word is ambiguous. A 0
with even index in ${\bf w}$ corresponds to an $a$ in ${\bf u}$,
while 0 with odd index in ${\bf w}$ corresponds to a $b$ in ${\bf
u}$; similarly, a 1 with even index in ${\bf w}$ corresponds to a
$d$ in ${\bf u}$, while 0 with odd index in ${\bf w}$ corresponds
to a $c$ in ${\bf u}$.

\begin{remark}\label{unique}
Let $v$ be an unambiguous  factor of ${\bf w}$. There is a unique
$v'\in \mathpzc F$ such that $v = f(v')$. Since $f$ is order
preserving, the least word of ${\mathpzc O}_w$ with prefix $v$ is
$f(\nu)$, where $\nu$ is the least word of ${\mathpzc O}_u$ with
prefix $v'$. By Theorem~\ref{Ou}, $\bf w$ is a suffix of $f(\nu)$.
\end{remark}

For $v\in\{0,1\}^*$, let $\sigma(v)$ denote the sum of the digits
of $v$.\begin{remark}\label{parity}  We note that
\begin{eqnarray*}
\sigma(f(g(x)))&=&\left\{\begin{array}{rl}0,&x\in\{a,d\}\\
1,&x\in\{b,c\}\end{array}\right.
\end{eqnarray*}
\end{remark}

\begin{lemma}\label{0000}Let $v$ be a factor of ${\bf w}$. If $|v|=4$ and $\sigma(v)$ is even, then $v$ is not ambiguous. In fact, let $pv$ be a prefix of ${\bf w}$. Then $|p|$ is odd; i.e., $v$ only appears in ${\bf w}$ with an odd index.
\end{lemma}
\pf Otherwise let $qxy$ be a prefix of ${\bf u}$ with $x$, $y\in
\{0,1\}$ and such that $f(g(xy))=v$. Write $pv=f(g(qxy))$, some
prefix $q$ of $\bf u$. Since the sum of the digits of $v$ is even,
$\sigma(f(g(x))\equiv \sigma(f(g(y))$ (mod 2). By
Remark~\ref{parity}, either $x,y\in\{a,d\}$ or $x,y\in\{b,c\}$.
This contradicts
Remark~\ref{index}.$\Box$\vspace{.1in}\vspace{.1in}

\begin{lemma}\label{0001*} Suppose that $v$ is an ambiguous factor of ${\bf w}$. Then $v$ is a factor of $(0001)^\omega$ or of $(1110)^\omega$.
\end{lemma}
\pf Every word of $\{0,1\}^3$ is a factor of at least one of
$(0001)^\omega$ and $(1110)^\omega$. There are exactly 8 words of
$\{0,1\}^4$ containing an odd number of 1's, namely,
0001,0010,0100,1000,1110,1101,1011 and 0111. By Lemma~\ref{0000},
if $|v|\ge 4$,  the de Bruijn graph of the length 4 factors of $v$
must be a subgraph of one of the cycles
$$0001\rightarrow 0010\rightarrow 0100\rightarrow 1000\rightarrow 0001$$
and
$$1110\rightarrow 1101 \rightarrow 1011\rightarrow 0111\rightarrow 1110.$$
The result follows.$\Box$\vspace{.1in}\vspace{.1in}
\begin{lemma} The words $0001000$, $0010001$, $0100010$, $1110111$, $1101110$ and $1011101$ are not ambiguous.\end{lemma}
\pf  Suppose that $0001000$ is ambiguous. Then $\bf w$ has a
prefix $p0001000$ where $|p|$ is even. Since $0000$ does not
appear in ${\bf w}$ with even index by Lemma~\ref{0000}, ${\bf w}$
must have a slightly longer prefix $p00010001$. Writing
$p=f(g(\mu))$, we find that ${\bf u}$ has a prefix $\mu abab.$
Since $a$ always has even index in ${\bf u}$, we find that ${\bf
u}$ has a prefix $\mu' aa$ where $g(\mu')=\mu$. This contradicts
Remark~\ref{index}.

A similar contradiction results if ${\bf w}$ has a prefix $pv$
where $v=0010001$ (resp., $0100010$) and $|p|$ is odd. In this
case ${\bf w}$ must have a prefix $p'0v$ where $p'0=p$. This
implies that $\bf u$ has a prefix $\mu abab$ (resp., $\mu acac$)
and therefore a prefix $\mu' aa$ (resp., $\mu' bb$) giving a
contradiction.

The above arguments, replacing $0$ with $1$, $a$ with $d$ and $b$
with $c$ show that $1110111$, $1101110$ and $1011101$ are not
ambiguous.$\Box$\vspace{.1in}

\begin{corollary}\label{ambig} No factor of $\bf w$ of length 8 is
ambiguous. In particular, every factor of ${\bf w}$ is a prefix of
an unambiguous factor of $\bf w$.
\end{corollary}
\pf By Lemma~\ref{0001*}, any ambiguous factor of $w$ is a factor
of one of $(0001)^\omega$ and $(1110)^\omega$. Every length 8
factor of those words contains one of the words shown not to be
ambiguous in the previous lemma.$\Box$\vspace{.1in}

To find least words in ${\mathpzc O}_u$, we reduced the problem of
finding a word with a long prefix to that of finding a word with a
shorter prefix. In the case of ${\mathpzc O}_w$, we can reverse
this procedure. Given a factor $x$ of $\bf w$, if $|x|\ge 8$, then
the least word of ${\mathpzc O}_w$ with prefix $x$ is given by
Remark~\ref{unique}. If $|x|<8$,  we find $v$, the least length 8
factor of $\bf w$ having $x$ as a prefix. The least word of
${\mathpzc O}_w$ having $x$ as a prefix also has $v$ as a prefix,
and is determined as per Remark~\ref{unique}.

\begin{theorem} Given a factor $x$ of $\bf w$, the least word of
${\mathpzc O}_w$ having $x$ as a prefix has the form $\mu{\bf w}$,
where word $\mu$ can be determined effectively from $x$.
\end{theorem}

\begin{theorem}
The least word of ${\mathpzc O}_w$ is $0{\bf w}$.
\end{theorem}
\pf We could make reference to length 8 factors of $\bf w$, but in
this special case it suffices to notice that 0000 is the least
length 4 factor of $\bf w$ and is unambiguous. Let $\lambda$ be
the least word of ${\mathpzc O}_w$. Word 0000 must be a prefix of
$\lambda$, and the only $v'\in {\mathpzc F}$ such that
$f(v')=0000$ is $v'=daba$. From Lemma~\ref{table}, we find that
the least word of ${\mathpzc O}_u$ with prefix $d$ is $\nu =d{\bf
u}$, which in fact has $daba$ as a prefix. It follows that $\nu$
is the least word of ${\mathpzc O}_u$ with prefix $v'$, so that,
as per Remark~\ref{unique}, $\lambda=f(\nu)=0{\bf
w}.\Box$\vspace{.1in}
\section{Acknowledgment}
Thanks to Jeffrey Shallit for making me aware of his elegant
conjecture.


\begin{thebibliography}{2}
\bibitem{AS} Jean-Paul Allouche and Jeffrey Shallit, {\em Automatic Sequences: Theory, Applications, Generalizations.} Cambridge University Press, 2003.
\bibitem{ARS} Jean-Paul Allouche, Narad Rampersad \& Jeffrey Shallit, Periodicity, repetitions, and orbits of an automatic sequence, {\em Theoret. Comput. Sci.} {\bf 410} (2009), 2795--2803.
\bibitem{lothaire} M. Lothaire, Combinatorics on Words,
Encyclopedia of Mathematics and its Applications 17,
Addison-Wesley, Reading, 1983.
\end{thebibliography}
 \end{document}